# Radii and Distances of Cepheids, I.
# Method and Measurement Errors

Martin Krockenberger, Dimitar D. Sasselov & Robert W. Noyes
17 September 1996

## Abstract

We develop a formulation of the Baade–Wesselink method which uses the Fourier coefficients of the observables. We derive an explicit, analytic expression to determine the mean radius from each Fourier order. The simplicity of this method allows us to derive the uncertainty in the mean radius due to measurement errors.

Using simulations and a recent dataset we demonstrate that the precision of the radius measurement with optical magnitudes is in most cases limited by the accuracy of the measurement of the phase difference between the light and the color index curve. In this case it is advantageous to determine the inverse radius, because it has normal errors.

## 1 Introduction

The Baade–Wesselink (BW) method of measuring radii and distances of Cepheid variable stars (Baade 1926, Wesselink 1946) is potentially very useful to establish the zero point of the Cepheid period–luminosity (PL) relation. Gautschy (1987) gives a good review of the status of the BW method. Gieren, Barnes & Moffett (1993) have used a sample of 100 Cepheids to derive the PL relation, which agrees within the uncertainties with the ZAMS-fitting distance scale. Laney & Stobie (1995) have used infrared photometry to derive a period–radius (PR) relation that is steeper than previous results. Pipepi et al. (1996) use an improved two color index method to find a much shallower PR relation. All the results mentioned above are based on different BW methods, photometric bandpasses and data sets. Laney & Stobie (1995) showed that the choice of photometric bandpass as well as the BW method affects the resulting PR relation.

This is the first paper of a series to improve the PR relation. In this paper we introduce our BW method and discuss the sources of measurement errors in BW radii. In subsequent papers we will apply our BW method to Cepheids and RR Lyr stars in the Galaxy and the LMC. Estimating the uncertainty in the radius of each individual Cepheid due to measurement errors is a problem in all realizations of the BW method. Both photometry and radial velocity measurements have errors that lead to an uncertainty in the BW radius. Knowing this uncertainty is important to be able to look for systematic errors. Good error estimates are also essential for determining the analytical form of the PR and the PL relation. Without an error estimate for each individual Cepheid radius or distance the quality of a PL or PR fit cannot be determined. Even though linear PR and PL relations are most widely used, there is still uncertainty whether or not this is the correct analytic form. Burki and Meylan (1986) find evidence for a break in the PR relation at a period of 10 days and Parsons (1972) fits a quadratic PR relation.

Balona (1977) includes the estimated measurement errors to derive maximum likelihood radii. The uncertainty Balona (1977) quotes for each stellar mean radius is the standard deviation of the individual radius determinations at each available photometric point. These errors include uncertainties in the phasing of photometry and radial velocities as well as possible problems with the assumption of a single valued color–surface brightness relation. The effects of the measurement errors cannot be separated from these other errors. Another problem with propagating measurement errors in the maximum likelihood method is the required interpolation of the velocity curve. This makes it difficult to assign velocity errors to the phase points used in the solution for the radius.

It is advantageous to use phase–integrated quantities to measure the BW radius because no interpolation is required and integrated quantities are less sensitive to noise. Caccin et al. (1981) introduced a BW method that determines the radius from an implicit, phase–integrated equation. In order to get phase integrated quantities they fit Fourier expansions to the data, which can easily be integrated. To estimate the influence of the measurement errors on the radius Caccin et al. (1981) did Monte Carlo simulations of several noise amplitudes. Using a certain $\zeta$ Gem data set they estimated the radius to be uncertain by about 5 %.

Another phase–integrated BW method uses the Fourier coefficients of light, color and velocity curves (Balona and Stobie 1979). Assuming a certain relation between the shape of the velocity curve and the shape of the color curve, the Fourier components of the light, color and velocity curves are used to determine the mean radius. The scatter in the resulting PR relation for 60 Cepheids using this Fourier method is larger than the scatter in the PR relation using a maximum likelihood method (Balona 1977). Thus the assumptions of this Fourier method might not be a good approximation for all Cepheids.

In this paper we present a new BW method using the Fourier expansions of the measured quantities to determine the radius. An advantage of this method is that measurement errors can easily be propagated. We go on and use this to discuss the influence of errors on the estimated radius.

## 2 The Fourier Baade–Wesselink Method

The radial pulsation of Cepheids provides a direct means of determining the absolute radius of each Cepheid via any of the existing variants of the classical BW method. The method is based on observing the radial motion of an emitting surface through the Doppler shift in its spectrum, as well as its flux and temperature variation. The radial velocities, when integrated appropriately over a certain time (or cycle), yield the linear displacements of the moving stellar surface. If the temperature were constant, the flux variation would reflect the change in radius of the emitting surface, and yield radius ratios at the same times. Combined, these two radius variations would lead to a solution for the mean radius.

However, as it pulsates a Cepheid changes its surface temperature. Both radius and temperature changes affect the total luminosity, which we observe as the Cepheid light curve. The effective temperature at the surface of any Cepheid is between 5000 and 7000 K; the variation due to the pulsation is usually about 1000 K. The emergent spectral distribution for that temperature range peaks in the optical region. As a result, the optical light variation is dominated by the surface brightness variation, while we need to determine the small variation caused by the change in radius. A common way to separate the surface brightness variation is to use the observed variation of the photometric color index as a proxy for the surface brightness. Then we have three observables: two light curves at two different wavelengths (bandpasses), and a velocity curve; we solve them simultaneously for the mean radius.

In this section we derive an equation that gives the mean radius of a pulsating star in terms of the Fourier coefficients of these three observables. This equation is very useful because it allows easy propagation of the measurement errors and gives insight into the main source of error in the radius. We assume that we have data with good coverage over phase, $\phi$, of the observables and that the period of the Cepheid is known. The starting point of our BW solution is the relation between magnitude, surface brightness and radius

$$M = S - 5\log\Big(\frac{R}{R_\odot}\Big) + \text{const.} \qquad (2.1)$$

We assume that the surface brightness can be written as a power series of the temperature observable, $T(\phi)$, $S(\phi) = c_0 + \sum_{j=1} c_j T^j(\phi)$. The temperature observable might be a color index, or a temperature sensitive line ratio. In this paper we define the temperature observable to be the B–V color index.

Then all observables, which are measured radial velocity V, magnitude M and color index T, can be Fourier expanded to give

$$X = X_0 + \sum_{k=1}^{m} X_k \cos(2\pi k(\phi + \phi_k^X)) \qquad (2.2)$$

where $\phi$ is the phase ranging from 0 to 1.

To write eq. (2.1) in terms of the Fourier coefficients of the observables we need to express the radius in terms of the velocity. We allow for a phase–dependent conversion factor (p–factor) from radial to pulsational velocity, $p(\phi)$, thus

$$R(\phi) = R_0 + \Delta r(\phi) = R_0 + P \int_{\phi_0}^{\phi} p(\phi')\Big(V(\phi') - \gamma\Big) d\phi' \qquad (2.3)$$

Here $R_0$ is the unknown mean radius, $P$ the period of the Cepheid and $\gamma$ the systemic velocity. We assume that $p(\phi)$ can also be Fourier expanded, so that we can write

$$\gamma = \frac{\int_0^1 p(\phi) V(\phi) d\phi}{\int_0^1 p(\phi) d\phi} = V_0 + \frac{1}{2p_0} \sum_{j=1} V_j p_j \cos(2\pi j(\phi_j^V - \phi_j^p)) \qquad (2.4)$$

Even though we set up the equations very generally, we will now solve them for the simplest case. This demonstrates the concept and keeps the algebra simple. Later we will discuss a more general solution. In the simplest case the p-factor is constant over phase and the color index–surface brightness relation is linear. Thus $p = p_0$, $\gamma = V_0$, $S = c_0 + c_1 T$ and

$$\Delta r(\phi) = \sum_{k=1}^{m} \frac{p_0 P V_k}{2\pi k} \sin(2\pi k(\phi + \phi_k^V)) \tag{2.5}$$

Substituting all Fourier expansions, equation (2.5) and the linear color index–surface brightness relation into eq. (2.1) we have the following equation.

$$M_0 + \sum_{k=1}^{m} M_k \cos(2\pi k(\phi + \phi_k^M)) = c_0 + c_1(T_0 + \sum_{k=1}^{m} T_k \cos(2\pi k(\phi + \phi_k^T)))$$
$$- 5\log\left(1 + \sum_{k=1}^{m} \frac{p_0 P V_k}{2\pi k R_0} \sin(2\pi k(\phi + \phi_k^V))\right) - 5\log\left(\frac{R_0}{R_\odot}\right) + \text{const.} \tag{2.6}$$

The key step in our BW formalism is to separate out the Fourier orders from each other. This we can do if we linearize the logarithmic term in eq. (2.6). We use $\ln(1 + x) \simeq x$, if x $\ll$ 1. In our case x = $\Delta r / R_0$, which is typically less than 0.2 for PopI Cepheids, thus the approximation might be acceptable for these stars. We will check the validity of this approximation below.

After approximating $\log(1 + x)$ by $x \log(e)$ we use the orthogonality of sin and cos to break eq. (2.6) up into a system of equations. We multiply eq. (2.6) by $\sin(2\pi j\phi)$, or $\cos(2\pi j\phi)$ and integrate over phase to find

$$M_j \sin(2\pi j\phi_j^M) = T_j \sin(2\pi j\phi_j^T) c_{1j} + \frac{5\log(e) p_0 P V_j \cos(2\pi j\phi_j^V)}{2\pi j R_{0j}}$$
$$M_j \cos(2\pi j\phi_j^M) = T_j \cos(2\pi j\phi_j^T) c_{1j} - \frac{5\log(e) p_0 P V_j \sin(2\pi j\phi_j^V)}{2\pi j R_{0j}} \tag{2.7}$$

Note that these equations are not a series of approximations. Each Fourier order contains all the information, thus in the absence of noise $R_{0j} = R_{0i}$ and $c_{1j} = c_{1i}$. Combining the two equations, the mean radius, $R_{0j}$ and the slope of the color–surface brightness relation, $c_{1j}$, are given by

$$R_{0j} = 0.3456 \; j^{-1} p_0 P \; \frac{V_j \cos(2\pi j(\phi_j^V - \phi_j^T))}{M_j \sin(2\pi j(\phi_j^M - \phi_j^T))}$$
$$c_{1j} = \frac{M_j \cos(2\pi j(\phi_j^V - \phi_j^M))}{T_j \cos(2\pi j(\phi_j^V - \phi_j^T))} \tag{2.8}$$

In the absence of noise, each Fourier order $j$ will give the same values of $R_{0j}$ and $c_{1j}$. For real Cepheid light, color and velocity curves the values of $R_{0j}$ and $c_{1j}$ will be affected by errors and thus be different for each Fourier orderx.

## 2.1 Effects of Linearization

The correct radius term of eq. (2.1) is given by

$$f_1(\phi) = 5\log\left(\frac{R_0 + \Delta r(\phi)}{R_\odot}\right) \tag{2.9}$$

This must be compared to the linearized radius term, given by

$$f_2(\phi) = 5\log(e)\frac{\Delta r(\phi)}{R_0} + 5\log\frac{R_0}{R_\odot} \tag{2.10}$$

We found that the Fourier amplitudes of typical Cepheid radius curves are reduced by the linearization (see below). This decrease is smallest for the first order and increasingly larger for the higher orders. The Fourier phases do not change at all between the linearized, $f_2$ and the full radius term, $f_1$.

The reduction in the amplitude of the radius curve leads to a smaller mean radius. This effect increases with large velocity amplitudes, the steepness of the piston phase and long periods (large $\Delta r$). The effect of the linearization on the determined radius can be removed by a correction factor given by the ratio of the Fourier amplitudes of the $f_{1,2}(\phi)$ curves.

The procedure for determining this correction factor is the following. We first solve eqs. (2.8) to get the mean radius, $R_0$. With this mean radius we determine $f_{1,2}$ and their respective Fourier coefficients. The ratio of the $i$th order Fourier amplitudes of $f_1$ and $f_2$ is the correction factor for the mean radius determined from the $i$th Fourier order.

Below we will examine the effectiveness of this procedure with simulations.

## 2.2 Measurement Errors

The simplicity of eq. (2.8) allows direct propagation of the measurement errors to estimate the uncertainty in the derived radius.

### 2.2.1 Errors in the Fourier Fit

For each of the observables $X$, where $X = V$, $M$ or $T$, the Fourier amplitudes, $X_n$, and Fourier phases, $\phi_n^X$, as well as their respective uncertainties are derived from a least squares fit of the velocity, magnitude and color index curves to Fourier expansions of the form

$$X = X_0 + \sum_{k=1}^{m} a_k \cos(2\pi k\phi) + b_k \sin(2\pi k\phi) \tag{2.11}$$

We assume that the period of the Cepheid is well known and that the velocity, magnitude and color index curves have been phased whith this period. The coefficients $X_k$ and $\phi_k^X$ of eq. (2.2) are related to $a_k$ and $b_k$ by

$$X_k = \sqrt{(a_k^2 + b_k^2)} \tag{2.12}$$

and

$$\phi_k^X = \frac{\arctan\left(\frac{-b_k}{a_k}\right)}{2\pi k} \tag{2.13}$$

The variances and covariances of $X_k$ and $\phi_k^X$ are given by

$$\sigma_{X_k X_l} = \frac{a_k a_l \sigma_{a_k a_l} + b_k b_l \sigma_{b_k b_l} + a_k b_l \sigma_{a_k b_l} + a_l b_k \sigma_{a_l b_k}}{\sqrt{(a_k^2 + b_k^2)(a_l^2 + b_l^2)}}$$

$$\sigma_{\phi_k^X \phi_l^X} = \frac{a_k a_l \sigma_{b_k b_l} + b_k b_l \sigma_{a_k a_l} - a_k b_l \sigma_{a_l b_k} - a_l b_k \sigma_{a_k b_l}}{4\pi^2 k\, l(a_k^2 + b_k^2)(a_l^2 + b_l^2)}$$

$$\sigma_{X_k \phi_l^X} = \frac{a_k a_l \sigma_{a_k b_l} - b_k b_l \sigma_{b_k a_l} - a_k b_l \sigma_{a_k a_k} + b_k a_l \sigma_{b_k b_l}}{2\pi l(a_l^2 + b_l^2)\sqrt{a_k^2 + b_k^2}} \tag{2.14}$$

The uncertainties, $\sigma_{a_k, b_k, a_l, b_l}$, are given by the covariance matrix of the least squares fit.

### 2.2.2 Errors in the Radius of each Fourier Order

The radius is determined from the coefficients $X_k$ and $\phi_k^X$ by eq. (2.8). The errors in the radius are given by $X_k$ and $\phi_k^X$ and their variances and covariances. We assume that there are no correlations between the Fourier coefficients of velocity, magnitude and color index. The variances and covariances of the mean radius are given by

$$\frac{\sigma_{R_{0i} R_{0j}}}{R_{0i} R_{0j}} = \frac{\sigma_{V_i V_j}}{V_i V_j} + \frac{\sigma_{M_i M_j}}{M_i M_j} + f_{r_v}(i) f_{r_v}(j) \sigma_{\phi_i^V \phi_j^V} + f_{r_m}(i) f_{r_m}(j) \sigma_{\phi_i^M \phi_j^M} + f_{r_t}(i) f_{r_t}(j) \sigma_{\phi_i^T \phi_j^T} +$$
$$\frac{f_{r_v}(i)}{V_j} \sigma_{V_j \phi_i^V} + \frac{f_{r_v}(j)}{V_i} \sigma_{V_i \phi_j^V} - \frac{f_{r_m}(i)}{M_j} \sigma_{M_j \phi_i^M} - \frac{f_{r_m}(j)}{M_i} \sigma_{M_i \phi_j^M} \tag{2.15}$$

with

$$f_{r_v}(j) = -2\pi j \tan(2\pi j(\phi_j^V - \phi_j^T))$$
$$f_{r_m}(j) = -2\pi j \cot(2\pi j(\phi_j^M - \phi_j^T))$$
$$f_{r_t}(j) = 2\pi j \frac{\cos(2\pi j(\phi_j^V - \phi_j^M))}{\sin(2\pi j(\phi_j^M - \phi_j^T)) \cos(2\pi j(\phi_j^V - \phi_j^T))} \tag{2.16}$$

Likewise the variances and covariances in the slope of the color–surface brightness relation, $c_1$, are given by

$$\frac{\sigma_{c_{1i} c_{1j}}}{c_{1i} c_{1j}} = \frac{\sigma_{T_i T_j}}{T_i T_j} + \frac{\sigma_{M_i M_j}}{M_i M_j} + f_{c_v}(i) f_{c_v}(j) \sigma_{\phi_i^V \phi_j^V} + f_{c_m}(i) f_{c_m}(j) \sigma_{\phi_i^M \phi_j^M} + f_{c_t}(i) f_{c_t}(j) \sigma_{\phi_i^T \phi_j^T} +$$
$$\frac{f_{r_t}(i)}{T_j} \sigma_{T_j \phi_i^T} + \frac{f_{r_t}(j)}{T_i} \sigma_{T_i \phi_j^T} - \frac{f_{r_m}(i)}{M_j} \sigma_{M_j \phi_i^M} - \frac{f_{r_m}(j)}{M_i} \sigma_{M_i \phi_j^M} \tag{2.17}$$

with

$$f_{c_v}(j) = 2\pi j \frac{\sin(2\pi j(\phi_j^M - \phi_j^T))}{\cos(2\pi j(\phi_j^V - \phi_j^T))\cos(2\pi j(\phi_j^V - \phi_j^M))}$$
$$f_{c_m}(j) = 2\pi j \tan(2\pi j(\phi_j^V - \phi_j^M))$$
$$f_{c_t}(j) = -2\pi j \tan(2\pi j(\phi_j^V - \phi_j^T)) \tag{2.18}$$

### 2.2.3 Do the Errors Follow a Normal Distribution ?

The radius, as given by eq. (2.8), is the ratio of two terms, $R_{0j} \propto x_j/y_j$.

$$x_j = V_j \cos(2\pi j(\phi_j^V - \phi_j^T))$$
$$y_j = M_j \sin(2\pi j(\phi_j^M - \phi_j^T)) \tag{2.19}$$

$x_j$ and its error is mostly determined by the velocity curve and $y_j$ and its error by the photometry. If the probability distribution functions of $x_j$ and $y_j$ are known, $p_{j1}(x_j)$ and $p_{j2}(y_j)$, then the probability distribution of $R_{0j}$ is given by

$$P_j(R_{0j}) = \int_{-\infty}^{\infty} p_{j1}(tR_{0j})p_{j2}(t)|t|dt \tag{2.20}$$

Below we will use simulations to demonstrate that $p_{j1}$ and $p_{j2}$ are well approximated by normal distributions. In this case the above integral can be written in terms of the error function, but still must be evaluated numerically.

To get some insight into the probability distribution of $R_{0j}$ we consider two limiting cases. If the error in $x_j$ is dominating the error in $R_{0j}$, the probability distribution of $R_{0j}$ is normal. Eq. (2.15) then gives a correct estimate of the width of the normal probability distribution. If the error in $y$ is dominating the error in $R_{0j}$, the probability distribution of $R_{0j}$ is not normal. But the inverse Radius, $R_{0j}^{-1}$ follows a normal distribution. In this case the error given in eq. (2.15) is not a good description of the true error. In this case we can instead use the inverse radius, $R_{0j}$, because its errors are normally distributed. The error of the inverse radius is easily calculated from eq. (2.15).

In Figure 1 we show the probablity distributions of the radius for a normally distributed inverse radius with a a mean of 0.01 and a $\sigma$ of 0.004. The probability distribution of the radius is asymmetric and the errors are different for positive and negative deviations.

### 2.2.4 Combining the Radii

The best mean radius is obtained by minimizing the likelihood function

$$L = \prod_j P_j(R_{0j} - R_0) \tag{2.21}$$

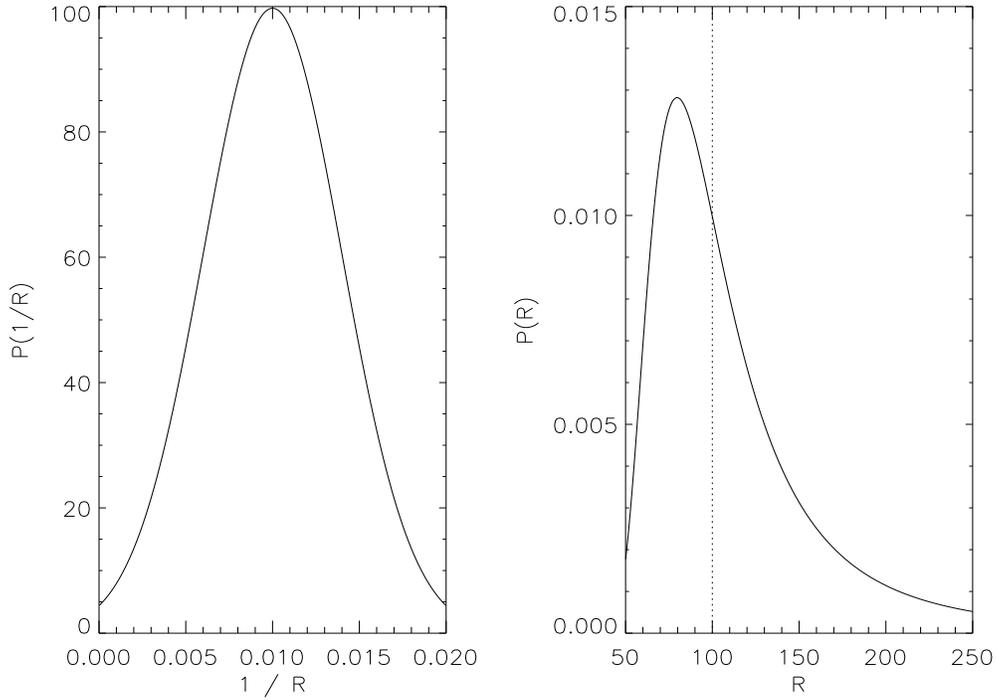

**Figure 1:** A Normal distribution of $R^{-1}$ leads to an asymmetric distribution in $R$. The distribution of radii peaks before the true radius and then there is a long tail of large radii.

In the case of normally distributed radius errors the final radius is calculated using

$$R_0 = \frac{\sum_j R_j \sigma^{-1}_{R_{0j}R_{0j}}}{\sum_j \sigma^{-1}_{R_{0j}R_{0j}}} \quad (2.22)$$

The error is given by

$$\sigma_{R_0 R_0} = \frac{\sum_{i,j} \sigma_{R_{0i}R_{0j}} \sigma^{-1}_{R_{0i}R_{0i}} \sigma^{-1}_{R_{0j}R_{0j}}}{\left(\sum_j \sigma^{-1}_{R_{0j}R_{0j}}\right)^2} \quad (2.23)$$

The same equations can be used for the inverse radius, if the probability distribution of the inverse radius is normal.

## 2.3 More General Solutions

In this chapter we deal with all the terms we omitted when deriving eq. (2.8).

### 2.3.1 Phase dependent p–factor

We now assume that we have the p–factor as a function of phase. The p-factor can be derived using models of Cepheid atmospheres (Sabbey et al. 1995) or using the measured widths and asymmetries of the spectral lines. With a phase dependent p–factor eq. (2.5) becomes

$$\Delta r = P\left(p_0 \sum_{j=1} \frac{V_j}{2\pi j} \sin(2\pi j(\phi + \phi_j^V)) + (V_0 - \gamma) \sum_{k=1} \frac{p_k}{2\pi k} \sin(2\pi k(\phi + \phi_k^p))\right.$$

$$+ \sum_{\substack{j,k=1 \\ j \neq k}} V_j p_k \left( \frac{\sin(2\pi((j-k)\phi + j\phi_j^V - k\phi_k^p))}{4\pi(j-k)} + \frac{\sin(2\pi((j+k)\phi + j\phi_j^V + k\phi_k^p))}{4\pi(j+k)} \right)$$

$$+ \sum_{j=1} V_j p_j \left( \frac{\sin(2\pi(2j\phi + j\phi_j^V + j\phi_j^p))}{8\pi j} \right) \Big) \tag{2.24}$$

The radius becomes in 1st order

$$R_0 = \frac{0.3456\, P}{M_1 \sin(2\pi(\phi_1^M - \phi_1^T))} \Big( p_0 V_1 \cos(2\pi(\phi_1^V - \phi_1^T)) + (V_0 - \gamma) p_1 \cos(2\pi(\phi_1^p - \phi_1^T)) +$$
$$\sum_{j=1} \frac{V_{j+1} p_j}{2} \cos(2\pi((j+1)\phi_{j+1}^V - j\phi_j^p - \phi_j^T)) +$$
$$\sum_{j=1} \frac{V_j p_{j+1}}{2} \cos(2\pi((j+1)\phi_{j+1}^p - j\phi_j^V - \phi_j^T)) \Big) \tag{2.25}$$

### 2.3.2 Non–linear Color Index–Surface Brightness Relation

Allowing for non–linear terms in the color index–surface brightness relation eq. (2.7) becomes

$$M_j \sin(2\pi j \phi_j^M) = \sum_{k=1} T_j^{(k)} \sin(2\pi j \phi_j^{T^{(k)}}) c_k + \frac{5\log(e) pP V_j \cos(2\pi j \phi_j^V)}{2\pi j R_0}$$
$$M_j \cos(2\pi j \phi_j^M) = \sum_{k=1} T_j^{(k)} \cos(2\pi j \phi_j^{T^{(k)}}) c_k - \frac{5\log(e) pP V_j \sin(2\pi j \phi_j^V)}{2\pi j R_0} \tag{2.26}$$

where $T_j^{(k)}$ and $\phi_j^{T^{(k)}}$ are the $j$th order Fourier components of $T^k$. In order to solve this system of equations we need to use higher Fourier orders. This requires high quality data, because if the higher orders contain little information, the solution will be dominated by noise. Using the first and second order Fourier components will allow to include terms up to $T^3$. This should be enough for most practical purposes.

## 3 Using Simulations to Asses the Effects of Errors

In order to demonstrate the effects of linearization and measurement errors on the radius we studied simulated data sets. The magnitude and velocity curves used for the simulations were derived from Fourier fits to the velocity and the magnitude curve of U Sgr, obtained by Bersier et al. (1994). For each phase point the Fourier fit allows us to "measure" the magnitude or the velocity. In order to exclude all systematic errors we assumed a mean radius and constructed a matching color index curve that solves eq. (2.1). We assumed either 20 or 40 points in both velocity and photometry. The phases of the data points were chosen randomly, but we required that there be no large gaps. The number of Fourier orders fitted for was determined from the largest phase gap, $\Delta_{max}$, in the data to ensure that the Fourier fit does not blow up between two data points. We chose the number of Fourier orders to be the smallest integer larger than $(\Delta_{max}^{-1} - 1)/2$. The typical number of Fourier orders fitted for was 3 – 5 for 20 data points and 6 – 8 for 40 data points.

An important quantity in determining BW radii is the phase difference, $\Delta\phi_j$, between the $j^{\text{th}}$ order Fourier phase of the color index, $T$, and magnitude, $M$, and its error, $\sigma_{\Delta\phi_j}$. This phase difference would be zero if the radius were constant. As the change in radius is small, $\Delta\phi_j$ also is small, because most of the change in brightness is mostly due to the temperature change for an optical bandpass. If infrared (IR) magnitudes and colors are used $\Delta\phi_j$ becomes larger.

Because $\Delta\phi_j$ is very small and in the denominator of eq. (2.8), the error in the radius is dominated in most cases by $\sigma_{\Delta\phi_j}$. A small change in $\Delta\phi_j$ leads to a large change in radius. Thus the inverse radius has normal errors, whereas the probablity distribution of the radius itself is asymmetric (Figure 1). Therefore the average inverse radius of 1000 simulations with typical photometric and velocity errors gives a very good estimate of the true inverse radius, but the average radius is significantly larger than the true value due to a few outliers.

The insights that we gained into the errors of BW radii are not restricted to our Fourier method. To demonstrate this we used a multidimentional minimization routine to solve eq. (2.6) simultaneously for all orders. The result of this procedure is a radius, derived from all Fourier orders, with no linearization. The radii derived using this method are very similar to the results of the Fourier method, but there is no information about the size and the behaviour of the errors.

In Figure 2 we show the distribution of inferred radii from 1000 simulated Cepheid data sets with typical measurement errors. We used a straight multidimentional minimization routine to derive the radii. The distribution of radii is asymmetric. Because the errors are dominated by $\Delta\phi_j$, the errors in $1/R_0$ are normal. The distribution functions are very similar to those in Figure 1.

### 3.1 Biases due to Measurement Errors

We used measurement errors with normal errors for which we give the width, $\sigma$, relative to the peak to peak amplitude of the Cepheid. The velocity errors we used have a $\sigma$ of 0, 5, 10 and 20 % of the amplitude of the velocity curve. For the photometric errors we used a $\sigma$ of 0, 1, 3 and 5 % of the amplitude of the light curve. We assumed that the error in the color index is 0.75 of the photometric error.

We simulated data of two different periods, 2 and 20 days, for which assumed a mean radius of $28.8 R_\odot$ and $97.6 R_\odot$ respectively. For a period of 2 days and our choice of velocity and magnitude curves $\Delta\phi_1 = 0.029$, and for 20 days $\Delta\phi_1 = 0.073$. $\Delta\phi_1$ becomes larger for the longer period, because $\Delta r/R_0$ becomes larger. The radius corrections due to linearization are small; the uncorrected first order radii are smaller by 0.4 % for a period of 2 days and 1.3 % for a period of 20 days.

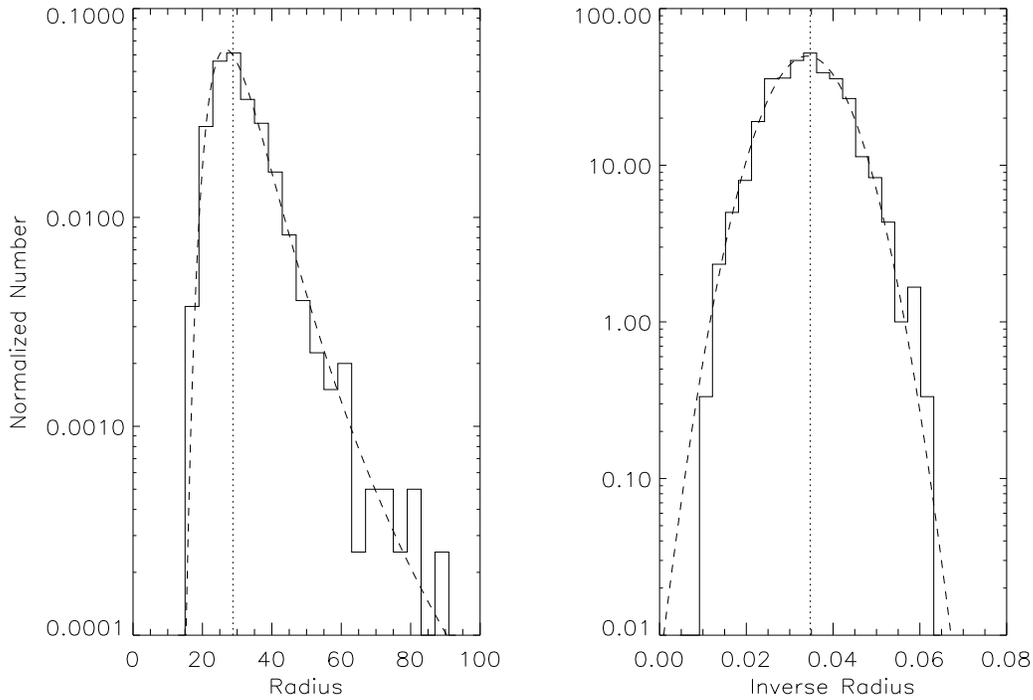

**Figure 2:** Distribution of the inferred radii from 1000 simulated Cepheid light, color and velocity curves. We used a multidimentional minimization routine, photometric errors of 3 % and velocity errors of 5 %. The inverse radii are well fitted by a Gaussian, indicating that the photometric errors are dominant. This is confirmed by applying the Fourier method to calculate the errors. The straight average of the radii is 10 % larger than the true radius.

In Figure 3 we compare the average radii from 1000 simulated data sets. We used a straight BW method and also the Fourier method. In the straight method we use a multidimentional minimization routine and a straight average of all 1000 radii. The mean radii of this method turn out to be significantly too large due to some outliers in the long tail of large radii. The Fourier method examines the errors and therefore averages the inverse radii for large photometric errors. The results of the Fourier method are very close to the true radius even for the largest photometric errors. We conclude that the straight BW radii are biased towards larger radii, especially for large photometric errors and short period stars. The Fourier method gives the same distribution of radii, but because the errors are understood, the mean radius is unbiased.

The bias towards larger radii due to photometric errors in the straight BW method is dependent on $\Delta\phi_j$ and thus period dependent. We expect that the radii of short period Cepheids in a large sample will on average be overestimated, leading to a larger offset and smaller slope in the PR–relation. One way to reduce this systematic bias is to use an infrared (IR) bandpass (Welch 1994, Laney and Stobie 1995) for the light curve. $\Delta\phi_j$ is larger if the magnitude is measured in the IR, because the IR magnitude of the Cepheid is more sensitive to the radius than to the temperature. We thus expect that the slope of the PR relation is steeper, if IR magnitudes are used. This is confirmed by the results of Laney & Stobie (1995).

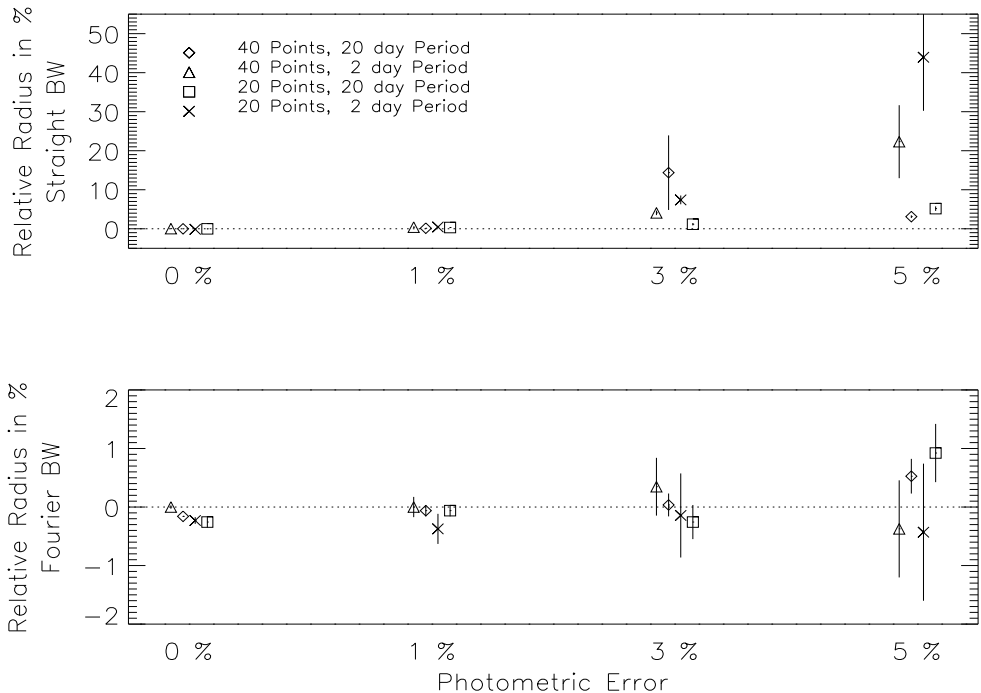

**Figure 3:** Mean radii for 0 % velocity error, given photometric error, period and number of points. The straight BW radii are calculated using a multidimensional minimization routine. Shown is the average over 1000 simulations. The error bars are the error in the mean. Note the change in scale from the straight BW radii to the Fourier radii.

### 3.2 Biases due to Truncation of the Fourier Series

The small bias in the Fourier radii for no measurement errors in Figure 3 is due to truncation of the Fourier series. This bias increases if there are fewer points available. If there are only very few points then the maximum and the minimum of the velocity curve are underestimated. Thus the Fourier amplitudes become smaller. For our particular choice of velocity and light curves the Fourier amplitudes of the velocity curve are more underestimated than the amplitudes of the light curve. Therefore the average radius is estimated to be 0.3 % smaller than the true radius. This effect is very small compared to the measurement errors. But it suggests that the Fourier method is biased for data sets with a very small number of phase points.

### 3.3 Biases due to Linearization

For periods shorter than 20 days the effects of linearizing the logarithm are small. They are easily hidden in the measurement errors. In order to demonstrate that our scheme to correct for the effects linearization works we simulated data sets with periods of 50 and 100 days. The radii for those periods are $159 R_\odot$ and $229 R_\odot$ and the ratios of $\Delta r/R_0$ are 0.17 and 0.23 respectively.

We simulated 1000 data sets for each period with both 20 and 40 phase points. There were no deviations from the true radius inconsistent with the uncertainty of the mean. The correction factor was typically 2.2 % for a period of 50 days and 3.5 % for a period of 100 days. The errors in the mean were much smaller than 1 % . We therefore conclude that our scheme to eliminate the effects of linerarization is successful.

# 4 The Fourier Method Applied to Real Data

To demonstrate our new method we applied it to a set of real Cepheid data. We chose the dataset of Bersier et al. (1994) because it provides measurement errors for each measurement. We measured the radii of the 19 Cepheids which had both photometry and velocities.

We used all photometric points except for those of quality zero (see Bersier et al. 1994). We derived the radii using the V magnitudes, the B–V color index and the radial velocities. The photometry is in the Geneva system, which is slightly different from the Johnson system, but has been used for BW solutions before (Burki & Meylan 1986). We assigned an error to each photometric point by the formula 0.02 / q where q is the quality of the photometry ranging from 1 to 4. We used the same relation for the quality of the color index, p, to get the errors of the color index. These formulae give errors for which the reduced $\chi^2$ of the fit is close to 1 for most objects. We used the velocity errors given by Bersier et al. (1994). For 3 of the stars, DX Gem, V465 Mon and W Sgr, we subtracted the velocity of the binary orbit before doing a Fourier fit.

The number of Fourier orders we fitted for is again given by the largest phase gap. Comparing our Fourier coefficients to those derived by Bersier et al. (1994) we found that they are in good agreement.

In Table 1 we show the number of data points per star and the quality of the Fourier fit. The Fourier method might not be adequate for any of those stars for which the reduced $\chi^2$ of any of the three Fourier fits is large. The method might also not work for those stars whose data is only sufficient to fit 3 or less Fourier orders. Often the Cepheid curves cannot be well represented with 3 or less Fourier orders. The fits for SV Vul are particularly bad because the period is variable. The last column gives the mean residuals from a linear color index–surface brightness relation. Typical photometric errors are about 1 % . Thus most of the stars with good data seem to be consistent with a linear color index–surface brightness relation.

The radii we found are shown in Table 2 and Figure 4. We used a constant p–factor of 1.36 (Burki et al. 1982). The radius is proportional to the p–factor (eq. (2.8)), thus the radii given in Table 2 can easily be converted to a different constant $p$–factor. The errors in radius are dominated by the uncertainty in the phase difference between color index and magnitude, $\Delta\phi_j$.

**Table 1:** Data and quality. $n_{phot}$ and $n_{vel}$ are the respective number of photometric and velocity data points. $f_{vel}$ and $f_{phot}$ are the number of Fourier orders fitted for. $\chi^2_{vel}$, $\chi^2_{temp}$ and $\chi^2_{mag}$ are the respective reduced $\chi^2$ values of the Fourier fits. The last column gives the mean deviation from a linear color index–surface brightness relation in magnitudes.

| Name | Period [d] | $n_{vel}$ | $f_{vel}$ | $\chi^2_{vel}$ | $n_{phot}$ | $f_{phot}$ | $\chi^2_{temp}$ | $\chi^2_{mag}$ | res [mag] |
|---|---|---|---|---|---|---|---|---|---|
| SW Tau | 1.58356 | 20 | 4 | 7.17 | 27 | 5 | 9.00 | 22.27 | 0.012 |
| EU Tau | 2.10248 | 41 | 6 | 1.74 | 24 | 5 | 1.49 | 4.18 | 0.003 |
| BB Gem | 2.30821 | 14 | 3 | 6.19 | 22 | 5 | 5.87 | 14.52 | 0.023 |
| DT Cyg | 2.49909 | 56 | 7 | 1.78 | 23 | 3 | 0.53 | 1.92 | 0.010 |
| BE Mon | 2.70551 | 18 | 5 | 6.41 | 18 | 5 | 0.81 | 1.24 | 0.016 |
| V465 Mon | 2.71300 | 16 | 2 | 2.00 | 18 | 5 | 1.50 | 1.02 | 0.000 |
| DX Gem | 3.13678 | 19 | 3 | 1.44 | 21 | 5 | 2.96 | 2.42 | 0.002 |
| SZ Tau | 3.14914 | 24 | 4 | 1.89 | 24 | 5 | 0.62 | 2.08 | 0.011 |
| ST Tau | 4.03430 | 30 | 5 | 1.48 | 17 | 3 | 1.70 | 5.43 | 0.007 |
| V508 Mon | 4.13361 | 17 | 3 | 2.48 | 20 | 3 | 3.21 | 0.68 | 0.011 |
| T Vul | 4.43545 | 116 | 9 | 2.42 | 24 | 4 | 0.74 | 1.54 | 0.009 |
| U Sgr | 6.74523 | 48 | 8 | 0.91 | 48 | 7 | 0.31 | 1.46 | 0.013 |
| V440 Per | 7.57250 | 64 | 6 | 1.46 | 34 | 5 | 0.47 | 0.78 | 0.008 |
| W Sgr | 7.59494 | 95 | 10 | 1.23 | 66 | 7 | 0.95 | 1.17 | 0.014 |
| S Nor | 9.75425 | 30 | 5 | 0.74 | 48 | 9 | 0.43 | 0.44 | 0.013 |
| Zeta Gem | 10.15000 | 65 | 5 | 1.21 | 27 | 5 | 0.66 | 2.27 | 0.014 |
| V340 Nor | 11.28871 | 35 | 6 | 0.71 | 50 | 10 | 3.54 | 1.13 | 0.006 |
| X Cyg | 16.38571 | 132 | 15 | 1.77 | 16 | 3 | 7.98 | 26.95 | 0.015 |
| SV Vul | 45.00450 | 84 | 11 | 68.85 | 14 | 2 | 29.33 | 38.99 | 0.040 |

### 4.1 Comparison of Radii

Only some of the radii in Table 2 are well determined. Some data sets are not good enough to measure a believable radius with the Fourier method. DT Cyg for example has a $\Delta\phi_1$ that seems too small for its period. We fitted its light and color index curves with 4 Fourier orders instead of 3 and found that the resulting reduced $\chi^2$ of the fit is almost unchanged. However, $\Delta\phi_1$ changes from $0.020 \pm 0.007$ to $0.034 \pm 0.010$ and the radius goes down from $66.9~^{+202.8}_{-28.8} R_\odot$ to $29.3~^{+31.3}_{-10.2} R_\odot$. This shows that there are not enough photometric data points for DT Cyg to determine the radius with the Fourier method.

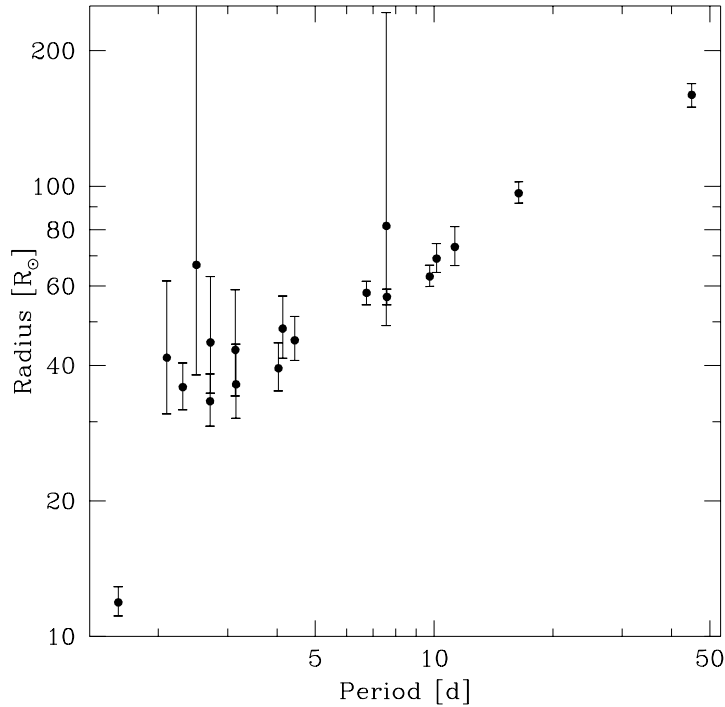

**Figure 4:** Period – Radius relation.

In Table 3 we compare the radii we have confidence in to previous radius determinations. The comparison shows that there are large differences in the various radii. The most overlap in data is between our radii and the radii by Ripepi et al. (1996), because both use the same velocity data.

It is interesting to note that the PR–relation of Ripepi et al. (1996) is much shallower than the PR–relation of Gieren et al. (1989) or Laney & Stobie (1995). We also find a very shallow relation. The data Ripepi et al. (1996) and we have in common are the velocity data of Bersier et al. (1994). We suspect that this set of velocities requires the usage of a different p–factor to be consistent with the velocities used by other groups.

As an example we examined the discrepancy between our result for T Vul and the result of Gieren, Barnes & Moffett (1989). We chose T Vul, because it shows the largest discrepancies. In order to exclude the effects of different BW realizations we applied our method to their data set. We used their B–V data and found $\Delta\phi_1 = 0.042 \pm 0.003$, in good agreement with our value. Thus the photometry is in agreement between the two data sets. The velocity curve of Gieren et al. (1989) is very noisy, leading to a first order Fourier amplitude of only 11.7 km/sec compared to 14.4 km/sec for the data set of Bersier et al. (1994). This is the reason why we find a smaller radius of $39.7^{+11.5}_{-7.5} R_\odot$ using the data set of Gieren et al. (1989). The discrepancy seems mostly due to noise in the velocity curve used by Gieren et al. (1989). The errors of the radii do overlap, thus they are not inconsistent. The 90 % confidence error bars on the Gieren et al. (1989) data is about 25 % . This partly explaines the large scatter

**Table 2:** Measurements of the mean radius of Cepheids, using a $p$ factor of 1.36. $\Delta\phi_1$ is the phase difference between color index and magnitude in the first Fourier order. $R_0$ is the corrected mean radius derived using eqs. (2.8). The errors in $R_0$ are the 90 % confidence interval. The other errors are 1 $\sigma$ errors. $c_1$ is the slope of the linear color index–surface brightness relation, obtained using eqs. (2.8).

| Name | Period [d] | $\Delta r/R_0$ | $\Delta\phi_1$ | $R_0$ [$R_\odot$] | $c_1$ |
|---:|---:|---:|---:|---:|---:|
| SW Tau | 1.58 | 0.06 | -0.042 +/- 0.002 | $11.9^{+1.0}_{-0.8}$ | 1.82 +/- 0.02 |
| EU Tau | 2.10 | 0.01 | -0.023 +/- 0.004 | $41.6^{+20.0}_{-10.4}$ | 1.84 +/- 0.05 |
| BB Gem | 2.31 | 0.04 | -0.025 +/- 0.002 | $35.8^{+4.7}_{-3.9}$ | 1.93 +/- 0.02 |
| DT Cyg | 2.50 | 0.01 | -0.020 +/- 0.007 | $66.9^{+202.8}_{-28.8}$ | 1.86 +/- 0.07 |
| BE Mon | 2.71 | 0.04 | -0.039 +/- 0.003 | $33.3^{+5.0}_{-4.0}$ | 1.75 +/- 0.03 |
| V465 Mon | 2.71 | 0.02 | -0.027 +/- 0.004 | $45.0^{+18.0}_{-10.3}$ | 1.84 +/- 0.07 |
| DX Gem | 3.14 | 0.02 | -0.032 +/- 0.005 | $43.3^{+15.6}_{-9.1}$ | 1.95 +/- 0.07 |
| SZ Tau | 3.15 | 0.03 | -0.038 +/- 0.004 | $36.3^{+8.3}_{-5.8}$ | 1.73 +/- 0.05 |
| ST Tau | 4.03 | 0.05 | -0.037 +/- 0.002 | $39.4^{+5.5}_{-4.3}$ | 1.78 +/- 0.03 |
| V508 Mon | 4.13 | 0.03 | -0.048 +/- 0.004 | $48.3^{+8.7}_{-6.8}$ | 1.77 +/- 0.05 |
| T Vul | 4.44 | 0.05 | -0.043 +/- 0.003 | $45.5^{+5.9}_{-4.5}$ | 1.65 +/- 0.03 |
| U Sgr | 6.75 | 0.06 | -0.047 +/- 0.001 | $58.0^{+3.5}_{-3.5}$ | 1.61 +/- 0.02 |
| V440 Per | 7.57 | 0.01 | -0.032 +/- 0.011 | $81.6^{+161.5}_{-32.6}$ | 1.68 +/- 0.15 |
| W Sgr | 7.59 | 0.07 | -0.047 +/- 0.001 | $56.8^{+2.3}_{-2.3}$ | 1.65 +/- 0.01 |
| S Nor | 9.75 | 0.06 | -0.058 +/- 0.001 | $63.0^{+3.8}_{-3.1}$ | 1.64 +/- 0.02 |
| Zeta Gem | 10.15 | 0.05 | -0.056 +/- 0.002 | $69.1^{+5.5}_{-4.8}$ | 1.57 +/- 0.03 |
| V340 Nor | 11.29 | 0.04 | -0.064 +/- 0.003 | $73.3^{+8.1}_{-6.6}$ | 1.66 +/- 0.04 |
| X Cyg | 16.39 | 0.13 | -0.068 +/- 0.002 | $96.5^{+5.8}_{-4.8}$ | 1.54 +/- 0.02 |
| SV Vul | 45.00 | 0.18 | -0.085 +/- 0.002 | $159.5^{+9.6}_{-9.6}$ | 1.81 +/- 0.03 |

in previous PR relations and the poor agreement among different measurements.

A comparison with the maximum likelihood radii of Laney & Stobie (1995) is difficult because we do not have access to the data set used. It seems though like there is fairly good agreement after allowing for 10 % error in the radii of Laney & Stobie (1995).

## 5 Conclusions

We have demonstrated a new method of determining the radii of pulsating stars and their errors due to measurement uncertainties. By applying of our method to the dataset by Bersier et al. (1994) and by simulations we found that the uncertainty in each individual radius measurement using optical magnitudes is dominated by the error in the phase difference between color index and magnitude. This phase difference is smaller for shorter period

**Table 3:** Cepheid radii from recent BW solutions. GBM are the radii from Gieren, Barnes & Moffet (1989), RBMR the radii from Ripepi et al. (1996), LS V are the V, B–V radii from Laney & Stobie (1995) and LS K are the K, J–K radii of Laney & Stobie (1995).

| Cepheid | Period [d] | this paper [$R_\odot$] | GBM [$R_\odot$] | RBMR [$R_\odot$] | LS V [$R_\odot$] | LS K [$R_\odot$] |
|---|---|---|---|---|---|---|
| T Vul    | 4.44 | $45.5^{+5.9}_{-4.5}$ | 38.2 | 48.8 |      |      |
| U Sgr    | 6.75 | $58.0^{+3.5}_{-3.5}$ | 57.3 | 34.5 | 63.3 | 50.5 |
| W Sgr    | 7.59 | $56.8^{+2.3}_{-2.3}$ | 63.3 | 58.6 |      |      |
| S Nor    | 9.75 | $63.0^{+3.8}_{-3.1}$ |      | 57.2 | 66.3 | 66.2 |
| Zeta Gem | 10.2 | $69.1^{+5.5}_{-4.8}$ | 64.9 | 86.2 |      |      |
| V340 Nor | 11.3 | $73.3^{+8.1}_{-6.6}$ |      | 71.6 | 66.8 | 74.4 |

Cepheids and leads to larger relative errors. This error is not caused by our Fourier method but affects any BW realization in the same way. Using the Fourier method we were able to identify this main source of measurement error.

It is important to determine the radius error for each individual star. Often the errors are very large, explaining why different radius measurements of the same star arrive at very different results. The PR–relation changes if the errors are ignored, because the individual errors are often asymmetric and very different in size.

If the measurement errors are known, the radius determination can be repeated on a large set of simulated data. The simulations will give the uncertainty and a possible bias due to the quality of the data and the method used. The radius determined from the data can then be corrected for this bias.

It should be possible to eliminate all biases due to the particular BW method used with simulated data sets. Systematic biases due to the quality of the data can also be eliminated with simulations. A problem that still needs to be solved before reliable BW radii can be obtained is the value and the phase dependence of the p–factor. This will be addressed in a subsequent paper in this series.

**Acknowledgments** We thank E. Feigelson and two anonymous referees for very helpful comments and suggestions.